\documentclass[12pt,preprint]{elsarticle}
\usepackage{graphicx}

\def\be{\begin{equation}}
\def\ee{\end{equation}}
\def\bea{\begin{eqnarray}}
\def\eea{\end{eqnarray}}

\begin{document}

\title{A new parallel simulation technique}

\author{Jose J. Blanco-Pillado}
\ead{jose@cosmos.phy.tufts.edu}

\author{Ken D. Olum\corref{cor1}}
\ead{kdo@cosmos.phy.tufts.edu}
\cortext[cor1]{Corresponding author}

\author{Benjamin Shlaer}
\ead{shlaer@cosmos.phy.tufts.edu}

\address{Institute of
  Cosmology, Department of Physics and Astronomy, Tufts University,
  Medford, MA 02155, USA.}

\begin{abstract}

We develop a ``semi-parallel'' simulation technique suggested by
Pretorius and Lehner, in which the simulation spacetime volume is
divided into a large number of small 4-volumes which have only initial
and final surfaces.  Thus there is no two-way communication between
processors, and the 4-volumes can be simulated independently without
the use of MPI.  This technique allows us to simulate much larger
volumes than we otherwise could, because we are not limited by total
memory size.  No processor time is lost waiting for other processors.

We compare a cosmic string simulation we developed using the
semi-parallel technique with our previous MPI-based code for several
test cases and find a factor of 2.6 improvement in the total amount of
processor time required to accomplish the same job for strings
evolving in the matter-dominated era.

\end{abstract}

\begin{keyword}

simulation, parallel processing, cosmic strings

\end{keyword}

\maketitle

\section{Introduction}

A common task in computational physics is the simulation of some
large physical system.  If the system is too large to be represented
on a single computer, or the resulting simulation would be very slow,
one simulates it using a number of processors (cores) working in
parallel.   Typically the spatial volume to be simulated is divided
into regions, and each processor handles one region.  At the
boundaries of the regions the processors must communicate using some
protocol such as MPI.

This technique has been used successfully for many simulations, but it
is not without drawbacks.  First, the processors must all be available
simultaneously.  In certain cases the number of processors must have
special properties, such as being a perfect cube.  In any case, the
largest domain that can be simulated is limited by the number of
available processors multiplied by the capability of each.  If one processor
crashes, due either to hardware failure or problems with the code, the
entire simulation must terminate.

Furthermore, for certain simulations there is a serious problem of
load balancing.  This poses no difficulty if the processors are
identical and the amount of work that each must do is the same.  For
example, to evolve equations of motion on a uniform grid one performs
the same operations on each point, so if all processors handle the
same number of points, there is no load balancing problem.  But for
other cases, the work is very different.  In the cosmic string example
to be discussed below, at late times many volumes are free of string
and incur no simulation overhead, while others have densely-packed
string points and require much work.  When the load is not balanced in
a conventional parallel simulation, all processors wait for the
slowest processor, and perhaps only a tiny fraction of the total
processing power can be utilized.

We describe here a different division of simulation work among
processors.  The idea is originally due to Pretorius and Lehner
\cite[\S 4.3]{Pretorius:2003wc}, although we discovered it
independently and did not learn of their work until later.  The
fundamental difference is that instead of dividing the spatial volume
to be simulated into as many regions as we have processors, we divide
the spacetime 4-volume to be simulated into a much larger number of
4-dimensional regions.  Each available processor will simulate many of
these regions.  We construct the regions in such a way that they take
in information through some initial surfaces and produce information
that is transmitted through some final surfaces, but there are no
surfaces with a two-way flow of information \cite{Pretorius:2003wc}.
In general relativity, such a region is called \emph{globally
  hyperbolic}.

Since there is no two-way information flow, there is no need for
multiple processors to be running simultaneously.  Instead, the
simulation of each region is a well-defined task that one processor
can perform alone, given only that it has the initial-surface
information provided by predecessor regions.  Thus even a single
processor could perform a simulation of arbitrary size without concern
about memory usage, by simulating the 4-dimensional regions one by
one.  The total simulation volume is thus not limited by the available
number of processors, although of course large simulations run on few
processors will take a great deal of time to
complete.

Furthermore, there is no issue of load balancing.  The processors run
independently, so no processor ever sits idle waiting for others.
Suppose one region takes much longer than others.  Its successors must
wait for it to finish, but other regions can be run simultaneously,
even if they are at later times but distant in space.  Eventually all
regions that are not successors (directly or indirectly) of the slow
region will be completed.  In this case the simulation is running on a
single processor only, but no other processors are tied up waiting for
the slow region to complete.  They can be used to run other
simulations or unrelated tasks.

In order for this technique to work, it must be possible to construct
regions with the proper causal structure.  For example, if one's
simulation consists of many nodes each of whose action at each moment
depends on the state of all other nodes at the immediately preceding
time, the division into regions cannot be done (except trivially by
having one node in each region).  But in the case of a simulation of
causal physics in spacetime, the propagation of information is
limited by the speed of light.  A null 3-surface divides spacetime
into a future and a past side; it is not possible for any information
to propagate across it in the reverse direction.  Thus if the regions
are divided by null surfaces, the necessary conditions apply.  In
other areas of research, the propagation speed might be the speed of
sound or some other limiting speed.  As long as such a speed exists,
the necessary division of spacetime can be made.\footnote{Of course if
  the speed of information is too fast, the resulting regions may be
  too small for efficient simulation.}

In the rest of this paper we discuss this ``semi-parallel'' simulation
technique in some detail, and describe a cosmic string simulation
using it.  In the next section we discuss the division of the
simulation 4-volume.  We discuss the choice of region size in
Sec.~\ref{sec:tuning} and the management of a simulation consisting of
very many regions in Sec.~\ref{sec:management}.  In
Sec.~\ref{sec:strings} we describe the cosmic string simulation that
we did using the semi-parallel technique, and in
Sec.~\ref{sec:results} we compare the performance of this technique
versus conventional parallelism.  We conclude and summarize in
Sec.~\ref{sec:conclusion}.

\section{Geometry}\label{sec:geometry}

\begin{figure}
\begin{center}
\includegraphics[width=3.5in]{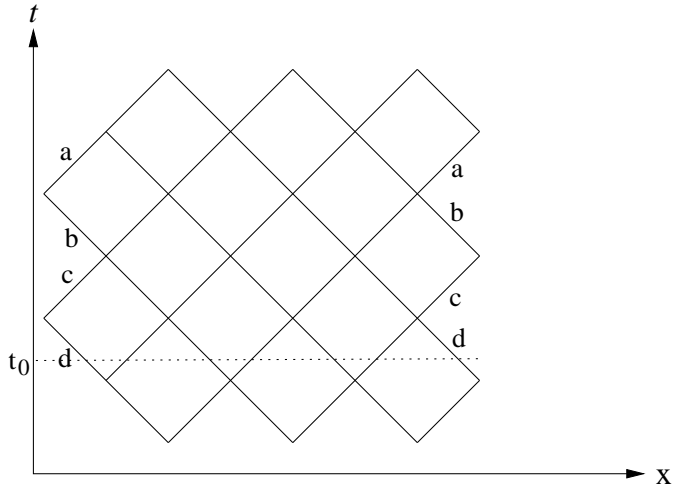}
\caption{A 2-dimensional spacetime volume divided into 15 regions by
  null lines.  Each square represents a region that can be simulated
  separately, using only conditions read in from predecessor regions.
  If the edges with matching letters are identified, our volume
  becomes periodic in the one spatial dimension.  The dotted line
  represents a possible initial time for a simulation using these
  regions.}
\label{fig:2d}
\end{center}
\end{figure}

In \cite{Pretorius:2003wc}, Pretorius and Lehner considered the
two-dimensional case and used square regions oriented with the time
direction diagonal, so that the edges would be null lines, as shown in
Fig.~\ref{fig:2d}.
We generalize this to four dimensions and use 4-cubes,
oriented so that the time direction lies along a main diagonal.  This
does not give null 3-surfaces as the boundaries, but we can fix the
problem by rescaling the time coordinate.

Let us work in units where the speed of light (or the maximum speed of
information flow) is 1.  We choose our regions to have unit edges and
consider a region whose past vertex is at the origin and whose future
vertex is thus at $(x,y,z,t) = (0,0,0,2)$.  The simulation regions
make up a 4-cubical lattice.  We can choose the four future-directed
generators of this lattice to be $(1/2,1/2,1/2,1/2)$,
$(1/2,-1/2,-1/2,1/2)$, $(-1/2,1/2,-1/2,1/2)$, and
$(-1/2,-1/2,1/2,1/2)$.  Then the four initial surfaces of our region
are 3-cubes that end at time $3/2$ at spatial positions
$(-1/2,-1/2,-1/2)$, $(-1/2,1/2,1/2)$, $(1/2,-1/2,1/2)$, and
$(1/2,1/2,-1/2)$.  The main diagonal of such a region has spatial
length $\sqrt{3}/2$ and temporal length $3/2$.  To make it null, we
can shrink our time coordinate by factor $\sqrt{3}$.

\begin{figure}
\begin{center}
\includegraphics[width=6.0in]{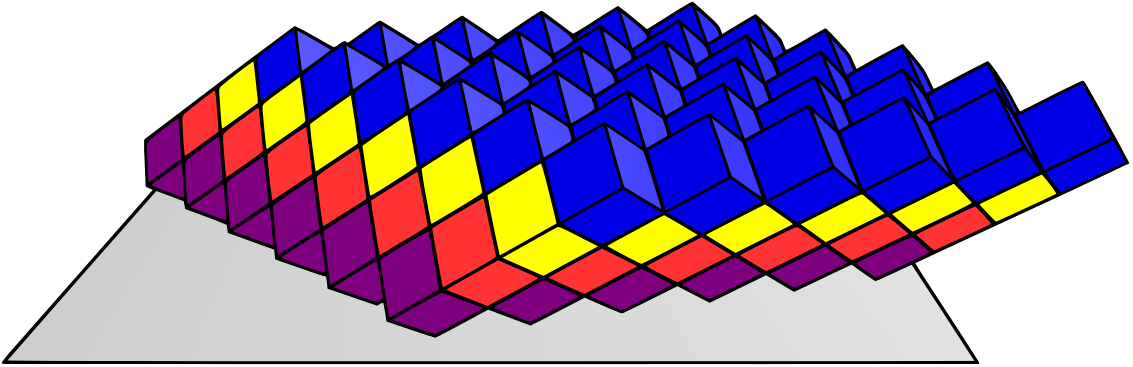}
\caption{The division of a 3-dimensional spacetime volume into cubical
  simulation regions.  Regions shown in the same color (shading) have
  the same starting time and touch each other along their edges.  A
  cube of one layer touches three cubes of the next layer along three
  of its faces.  The top points of the cubes of the bottom layer shown
  are the same as the bottom points of the cubes of the top layer
  shown.  In four dimensions, the volumes are 4-cubes.  A 4-cube of
  one layer touches four 4-cubes of the next layer along four of its
  eight 3-cubical faces.  The top points of the cubes of one layer are
  the same as the bottom points of the cubes four layers later.}
\label{fig:layers}
\end{center}
\end{figure}

If we follow a generator into the future and then another generator
into the past, we get a new region whose starting time is
the same as that of the original region.  Thus the regions can be
arranged into ``layers'' that share a common starting time.  The
analogous division of a 3-dimensional spacetime volume is shown in
Fig.~\ref{fig:layers}.

\subsection{Initial conditions}

\begin{figure}
\begin{center}
\vspace{-30pt}
\includegraphics[width=6.0in]{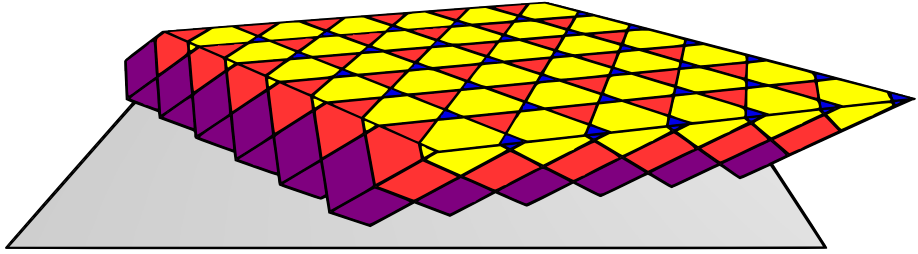}
\vspace{-30pt}
\caption{A slice through the simulation volumes shown in
  Fig.~\ref{fig:layers} at a time $t_0$.  The large red (lighter gray)
  triangles are near the latest points of cubes with no predecessors
  before $t_0$.  Initial conditions there can be generated
  immediately.  The hexagonal yellow (white) regions are in the next
  layer of cubes and the small triangular blue (darker gray) regions
  are in the final layer shown in Fig.~\ref{fig:layers}, which is the
  last for which initial conditions are necessary.  In four dimensions
  there are four layers of cubes that require initial
  conditions.}
\label{fig:slice}
\end{center}
\end{figure}

We will start our simulation at some particular initial time $t_0$.
The initial time surface will cut through the 4-lattice of simulation
regions.  The 3-dimensional analogue is shown in Fig.~\ref{fig:slice},
and the 2-dimensional analogue in Fig.~\ref{fig:2d}.  There will be
some cubes whose final vertex lies after $t_0$, but all of whose other
vertices lie at or before $t_0$.  Such a cube has no predecessors.  We
can generate initial conditions inside this cube, and then evolve them
until our ending point, writing out information on the four final
surfaces to be used by our successors.  All such cubes can be done
independently; they have no need to communicate with each other.

There will be another family of cubes that are cut further in their
pasts by the initial surface.  These require some conditions on the
$t=t_0$ hypersurface but also input from the first layer of initial
regions.  In all there will be four different families of cubes that
require initial conditions, one of which may be trivial if the cubes
are chosen to have vertices at $t_0$.  After these initial regions,
all subsequent regions will take their starting conditions only from
their predecessors.

\subsection{Boundary conditions}\label{sec:boundaries}

Often, as in our cosmic string simulation below, one wants to simulate
a finite volume with periodic boundary conditions.  Such conditions
can be implemented without additional effort merely by adjusting the
connections between simulation regions and their successors.  First,
suppose that we want only one region in each layer.  The four
successors of this region will all then be the unique region in the
subsequent layer, and so on.  When we finish simulating a region, we
write out four files corresponding to the four future surfaces.  When we go
on to the next region, we read these four files and consider them the
communication from our four virtual predecessors.

\begin{figure}
\begin{center}
\vspace{-0.8in}
\hbox{\raisebox{-1in}{\hspace{-0.5in}\includegraphics[angle=90,width=3.5in]{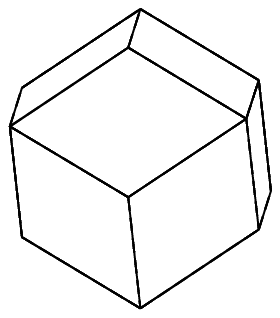}}
\includegraphics[width=2.0in]{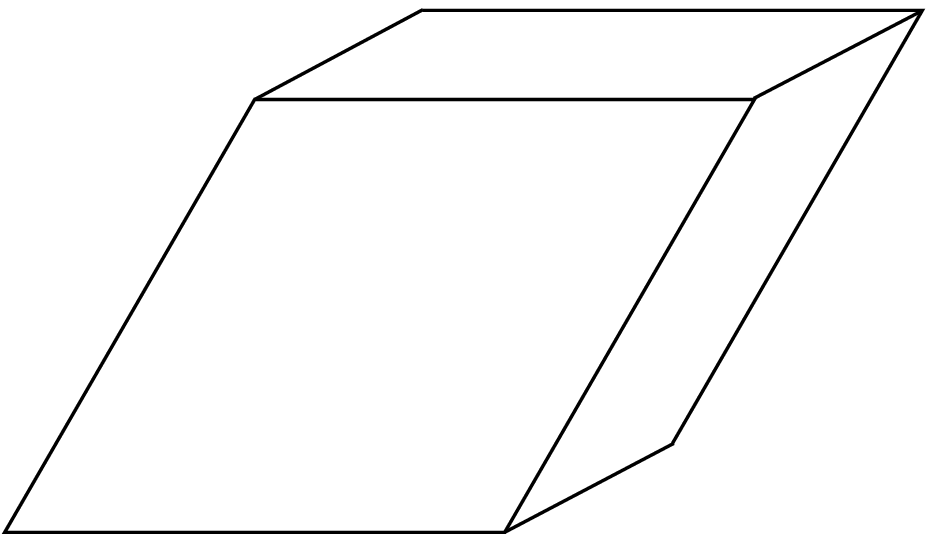}}
\vspace{-0.8in}
\caption{The most compact representation of the simulation volume is a
  rhombic dodecahedron, shown on the left.  But for most purposes it is
  easier to consider it a rhombohedron made of $60^\circ$ rhombuses,
  shown on the right.}
\label{fig:rhombohedron}
\end{center}
\end{figure}

A generic surface of constant time then intersects four regions, one from
each of the four layers active at that time.  The periodicity vectors of
the compactification can be found by going from a region to a
successor in one of the four future directions and then to a predecessor
in a different past direction.  Thus there are 12 periodicity vectors.
With the generators above, these are the 3-vectors having one
component 0 and the others each $\pm1$.  A point and its images form a
face-centered cubic lattice.  If we take as the representative of each
point the one which is closest to the origin, we see that the
simulation volume is a rhombic dodecahedron.  But
it is easier to understand if we reorganize the volume into a
rhombohedron made of rhombuses whose acute angles are $60^\circ$.
Both regions are shown in Fig.~\ref{fig:rhombohedron}.
With the coordinates we have been using, the edges of the rhombohedron
have length 1, and the spatial volume of the simulation region is
$1/\sqrt{2}$.

Suppose information travels at some speed $v$ through a periodic
simulation volume.  The volume influenced by the initial conditions at
a given point at time $t$ fills a sphere of radius $vt$.  When the
sphere around a point touches that around an image of the point, then
the existence of periodicity might affect the simulation results.  We
would like to delay this time as much as possible for a given
simulation volume, and thus we would like the spheres around a point
and its images to be close-packed.  Conveniently, the f.c.c. lattice
is a close-packed system, so we get this advantage for free.

\begin{figure}
\hbox{\includegraphics[width=3.0in]{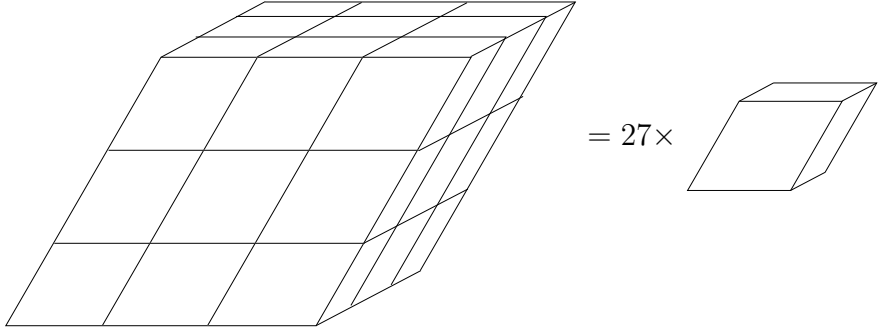}
\vbox to 2.0in{\vfill \hbox{$= 27 \times$} \vfill}
\vbox to 2.0in{\vfill \includegraphics[width=1.0in]{rhombohedron.eps} \vfill}}
\caption{A larger rhombohedral volume made of 27 rhombohedra.}
\label{fig:rhombohedron-cut}
\end{figure}

To simulate larger volumes, the easiest plan is to combine unit
rhombohedral volumes into a 3-dimensional array of such volumes, as
shown in Fig.~\ref{fig:rhombohedron-cut}.
One could choose the three dimensions of this
array independently, but in our case we chose them to be the same, so
the overall volume is again a rhombohedron, and the close-packing
property above is preserved.  Thus each layer consists of $N^3$
regions, for some ``split factor'' $N$.  Our largest simulations to date were
done with $N = 50$.

\section{Tuning}\label{sec:tuning}
How should we choose the size of our regions?  Clearly, the smaller the
regions the larger the ratio of their 3-surface area to their interior
4-volume, and thus the more time will be spent reading information
from predecessors and sending to successors as compared to evolution.
So, all other things being equal, larger regions will be more
efficient. However, there are two factors that argue for smaller
regions.  

First, the available memory per processor (core) may limit the amount
of simulation volume that can be handled by a single processor.  By
making the simulation volumes smaller than this limit, we buy the
opportunity to trade running time for space usage in the simulation
code.  The cost is an increase in the relative communication time, but
in our case we found that optimizing time at the expense of space in
our code was much more important than the increase in the small
percentage of time spent on communication.

Second, there is more opportunity for parallelism if one splits up the
simulation volume into more pieces.  The maximum number of processors
that can be running simultaneously is given by the number of regions
in each layer.  With perfect load-balancing the processors will all
advance simultaneously to the next layer, but when that cannot be
achieved, the realistic parallelism will be significantly less.  So,
for example, if you have one thousand processors available, you should
split up each layer into several thousand regions.

Even finer splitting may be desirable if some regions are more
difficult to simulate than others.  In this case, when the simulation
reaches the difficult regions, the number of processors that can be
used simultaneously will decrease. (Other processors are not tied up,
but still the available computational resources are not being used to
complete the simulation in the shortest possible time.)  A finer
split will give significantly more simultaneously-runnable regions
than there are processors, so if the possible parallelism is reduced
by difficult regions, more of the available processors will
nevertheless be kept busy.

The problem of difficult regions could also be solved by dynamical
splitting \cite{Pretorius:2003wc}.  A single region could be
subdivided into several regions to be handled separately, and the
initial data for the original region parceled out to different
processors.  Some of these subregions could be simulated in parallel
with others, so the difficult job is spread across several processors.
Once the difficult region is past, the final data from several regions
could be combined into initial data for a subsequent, larger region.
However, we have not implemented such a technique.

\section{Simulation management}\label{sec:management}

A typical parallel simulation involves a fixed number of processors,
each of which has a particular task, typically the simulation of a
specified region.  Using our method, there is a pool of a large number
of regions to be simulated, some of which are ready to go but most of
which are waiting for predecessors to be done.  To simulate these
regions we have a pool of available processors.  In principle one
could merely submit one job for each region to a batch-job scheduling
system, with a set of dependency conditions.  The scheduler would then
schedule each region on the next available processor.  However, we
have found that the individual regions often execute in a much shorter
period of time (e.g., 1 second) than the time it takes for batch
systems to schedule a new job (e.g., 30 seconds) , so this procedure
is quite inefficient.

Instead, we have found it useful to have a ``manager'' process that
assigns the regions to a variable-size pool of ``worker'' processes.
When a worker completes a region, it informs the manager that that
region has been completed and the manager replies with a request
giving the worker the next region on which to work.  If a worker
fails, the manager shuts down the entire simulation once the currently
running regions are finished.  We can repair the problem that caused
the failure and continue the simulation to do the remaining regions.
Each transaction between the worker and the manager is a single
exchange of network packets, so network overhead is minimal.

When some regions are more difficult than others, we can find
ourselves in the situation where there are more available workers than
regions that can be simulated in parallel.  When this occurs, the
manager first tells idle workers to sleep for a time comparable to the
time it takes to schedule a batch job.  If, during this time, another
worker finishes a region which makes more than one new region ready to
simulate, the manager wakes sleeping workers to simulate the newly
ready regions.  But if the situation persists, the manager tells the
sleeping workers to exit.  If at a later time there are persistently
more ready regions than running workers, the manager submits new batch
jobs to increase the number of workers.

The manager, like the simulation described below, was written in LISP.

\section{Semi-parallel cosmic string simulation}\label{sec:strings}

We have developed a large cosmic string simulation using this
technique to parallelize an algorithm developed earlier by Olum,
Vanchurin, and Vilenkin
\cite{Vanchurin:2005yb,Vanchurin:2005pa,Olum:2006ix}.  Our simulation
runs either in flat spacetime or in a flat Robertson-Walker universe.
In the latter case, we use comoving coordinates and conformal times,
so the causal structure is still that of Minkowski space.

\def\bx{\mathbf{x}}
\def\ba{\mathbf{a}}
\def\bb{\mathbf{b}}

Cosmic strings are astrophysically long, microphysically thin objects
which may have been formed early in the universe through a phase
transition \cite{Kibble:1976sj} or at the end of inflation driven by
superstring theory \cite{Sarangi:2002yt,Copeland:2003bj,Dvali:2003zj}.
See \cite{AlexBook} for further information.  In usual models, cosmic
strings cannot end, so strings form a ``network'' of infinite strings
and closed loops.  The string network could potentially be observed in
many ways such as cosmic microwave background variations
\cite{Kaiser:1984iv}, gravitational lensing
\cite{Vilenkin:1984ea,Gott:1984ef}, cosmic rays
\cite{Bhattacharjee:1989vu,Damour:1996pv,Vachaspati:2009kq},
gravitational waves \cite{Vachaspati:1984gt,Damour:2000wa} or early
reionization \cite{Olum:2006at}.

To compute any of these observational effects quantitatively requires
a knowledge of the total amount of string, the distribution of loops
and the spectrum of excitations on the strings.  The evolution of the
string network has so far resisted a complete analytic
description\footnote{Although some progress has been made in this
  regard.  See
  \cite{Austin:1993rg,Polchinski:2006ee,Dubath:2007mf,Polchinski:2007rg,Copeland:2009dk}},
so we simulate its evolution to determine the parameters needed to
predict observations.

Because the thickness of a cosmic string is so much less than the
sizes of structures that we expect to find on it, we can treat it
as an infinitely thin relativistic string.  If we ignore gravitational
effects (as we usually do in simulations), the mass density of the
string does not affect its evolution.  Thus one does not need
different simulations for different possible string energy scales.

When two strings cross, they switch partners with some probability
$p$.  For strings formed from a symmetry breaking transition in field
theory, $p$ is essentially 1, but for superstrings, $p$ can be
anywhere from 1 down to $10^{-3}$ \cite{Polchinski:1988cn}.  To
perform a simulation one must track the positions and motions of a
large network of strings, detect the intersections, and perform the
switching of partners.

The position of a cosmic string over time can be written $\bx(\sigma,
t)$, where $\sigma$ parameterizes the position on the string and $t$
is the usual time variable.  In the absence of spacetime curvature or
reconnections, the string obeys the Nambu-Goto equations of motion,
which can be written
\be
\ddot\bx(\sigma, t) = \bx''(\sigma, t)\,,
\ee
where a prime denotes differentiation with respect to $\sigma$ and a
dot differentiation with respect to $t$.  The solution is simply
\be
\bx(\sigma, t) = (1/2)[\ba(t-\sigma)+\bb(t+\sigma)]\,.
\ee
We use a piecewise linear form for the functions $\ba$ and $\bb$.

\begin{figure}
\begin{center}
\includegraphics[width=3.5in]{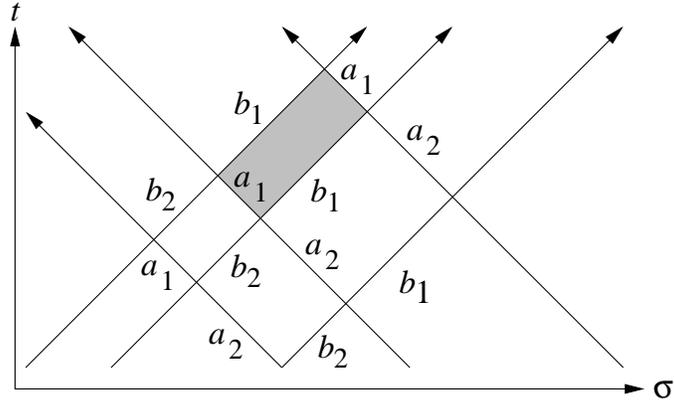}
\end{center}
\caption{A piece of the world sheet of a simulated string.  The
  functions $\ba$ is composed of linear pieces $\ba_1, \ba_2, \ldots$
  and similarly for $\bb$.  The shaded section shows the piece of the
  world sheet where $\ba_1$ is added to $\bb_1$ to give $\bx(\sigma,t)$.}
\label{fig:diamonds}
\end{figure}

\begin{figure}
\begin{center}
\includegraphics[width=3.5in]{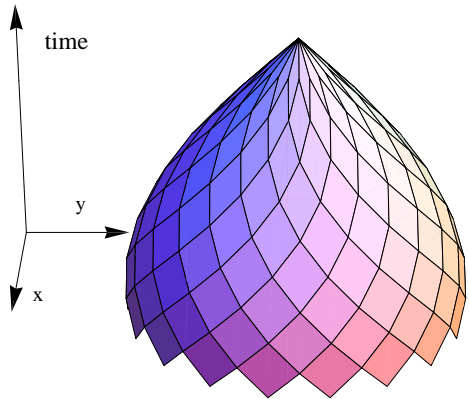}
\caption{The world sheet of a collapsing circle of cosmic string
  (approximated by a piecewise-linear shape) shown embedded in
  3-dimensional spacetime.}
\end{center}
\label{fig:worldsheet}
\end{figure}

The evolution of the string can be described by a 2-dimensional world
sheet embedded in 4-dimensional spacetime.  This surface is composed
of diamond-shaped regions where a given piece of $\ba$ is combined
with a given piece of $\bb$, as shown in Figs.~\ref{fig:diamonds} and
\ref{fig:worldsheet}.

In flat spacetime, each diamond is a piece of a plane in 4-space.
In the expanding universe, we implement expansion at first order in
each diamond, so that the future edges of the diamond are still
linear, but not parallel to the past edges, and the diamond is curved.
The main task of the simulation is to construct all the diamonds that
make up the world sheet, look for intersections between different
pieces of string, and reconnect the strings at the intersections.

This procedure has no ``simulation resolution'', in the sense that it
is able to simulate arbitrarily small segments of string.  In flat
space, it simulates the evolution exactly (to the limits of computer
arithmetic), given the piecewise-linear initial conditions.  For
details on the algorithm see
\cite{Vanchurin:2005yb,Vanchurin:2005pa,Olum:2006ix}.

We have parallelized our algorithm according to the procedure
described above.  The simulation 4-volume is divided into a large
number of regions in the shape of squished 4-cubes standing on one
corner.  Each diamond is created during the simulation of the region
that contains its earliest point.  At the end of simulating a region,
the processor passes on to its successors the information about all
world sheet diamonds that overlap each final surface.

Because diamonds are extended objects, they can stretch through
several simulation regions.  In fact, a diamond created in one region
may need to go to several successors.  In such a case, we deliver the
information about the diamond to each of these, along with a notation
of which portions of the diamond each successor should handle.
Each successor will pass the diamond on to its successors, until
eventually it reaches a single common successor, whereupon the pieces
from the predecessors are reassembled into a single diamond.

As the string network evolves, cosmic strings can intersect with
themselves, producing loops.  These loops may self-intersect further,
but eventually one finds some non-self-intersecting loops in periodic
trajectories.  These loops are the main source of potentially
observable effects today, so we are interested in the numbers of loops
of different sizes.  Small, non-self-intersecting loops will only
rarely rejoin the long-string network, so once such a loop is formed,
its further evolution is not usually of interest.  To speed the
simulation we record and remove such loops.

As a result of the removal of small loops, the total amount of string
in the network declines with time.  The typical distance between the
remaining strings at time $t$ grows as roughly $0.15 t$.  In a large
simulation, this distance eventually becomes larger than the size of
each simulation region, and therefore most regions contain no string
at all.  When we find a region with no strings in its initial
conditions, we do not need to simulate it at all: clearly there will
be no string on its output surfaces, and we must merely take note of
this fact.

\section{Results}\label{sec:results}

The results of our simulation from the point of view of cosmic string
physics will be presented elsewhere \cite{BPOS1}.  Here we discuss this
simulation from a computational point of view.

The largest simulations we have done involved periodicity length 2000
in units of the initial average distance between strings.  The total
time simulated was 3500 in the same units\footnote{Information in a
  string network flows only along the strings, resulting in a net
  speed of information flow through the volume about half the speed of
  light \cite{Olum:2006ix}.  This makes possible a simulation of
  duration 2000 in a volume with periodicity distance 2000.  But we
  run for longer than that to monitor the evolution of loops to see if
  they are in non-self-intersecting trajectories.}.  We divided the
simulation volume according to the procedure of
Sec.~\ref{sec:boundaries} with $N=50$, so that each region has size 40
and volume $40^3/\sqrt{2}$.  (``Size'' here is the edge length of the
rhombohedral region.)  The $N=50$ division gives 125,000 regions in
each layer, and about 54 million regions in all.  The job was done on
400 processors in elapsed time about 66 hours.  Of the 54 million
regions, only about 7 million contained string to be simulated.  The
others were empty.

In such a simulation the number of individual pieces of string grows
at early times because segments are divided by reconnections, but then
shrinks at late times because loops are removed.  The peak number of
segments in our size-2000 simulation reached about 14 billion.  That
makes this among the largest simulations ever performed if one counts
the largest total amount of data existing at one moment of simulation
time.  For comparison, the Millennium Simulation \cite{Springel:2005nw}
used just over 10 billion particles.  In the course of our simulation,
about 1 trillion diamonds were created and about 10 billion loops of
cosmic string produced.

The total comoving length of string in the simulation at a given time
does not depend strongly on the cosmological model.  However, because
strings are (typically) not straight, the stretching of a given
segment of string is less than the overall expansion of the universe,
and so its length in comoving units decreases.  Thus the total number
of segments (and consequently the simulation effort) required to
simulate a certain comoving length of string at a late time is much
higher in expanding-universe simulations.  Expansion in the matter era
is faster than that in the radiation era, so the matter era
simulations feel this issue most strongly.

When the number of regions in each layer is much larger than the
number of available processors, and the difference between one region
and the next is not too large, the processors can all run
simultaneously.  The size 2000 run described above was done in flat
space.  In this case, there were always thousands of regions that were
ready to go any given time, except when the simulation was nearly
finished.  So the 400 processors that we used were always busy.

However, in a matter era run the situation is very different.  After
the beginning stages of the simulation are past, often only a few
processors are running.  These are the ones that have most of the
string, which is in quite small segments.  Meanwhile, all regions that
are not in the future of these regions have been completed, so there
is nothing more to do.  In this case the elapsed time depends on the
simulation of regions with the largest number of string segments, but
the rest of the available processors are not tied up waiting, so they
can be used for other things.  In the same simulation with
conventional parallelism, most of the CPU time would be wasted as the
processors wait for the processor with the highest number of string
segments to run.

The difference between cosmological models also affects the choice of
the optimal region size.  Memory constraints limit us to a region size
of order 65 per processor, meaning that total periodicity distance
2000 must be split by a factor of at least 31 (in each direction).
The communications overhead in this simulation is generally
negligible, so additional splitting has a low cost.

For matter-era simulations, a much more important issue is to split
the regions with the largest density of string at late times.  Thus in
that case we usually split into much smaller pieces whose size is of
order 17.\footnote{The initial conditions in our simulation are
  generated by the algorithm of Vachaspati and Vilenkin
  \cite{Vachaspati:1984dz}, which is non-local in that the initial
  strings in a particular region of spacetime depend on the generation
  of initial data outside that region.  This sets a minimum size for
  each simulation region.  If regions are smaller than $12\sqrt{2}
  \approx 16.97$ it is not possible for different regions in the first
  layer to generate consistent initial data, since they are not in
  communication with each other.}

Because we had our simulation based on the above algorithm already
running under MPI before we converted it to the semi-parallel
technique, we were able to compare the two techniques.  Unfortunately,
along the way we also did quite a bit of optimization of the code, so
the total CPU times for the new code are significantly lower and the
comparison is not as direct as one might wish.  Some of this
improvement was made possible because the new technique avoids memory
limitations and so permitted us to use algorithms that were faster
(mostly because of additional caching) at the expense of greater
memory use.  But the majority of the improvement was simply careful
optimization of time-intensive portions of the code and is not
related to the change of parallelization techniques.

In Table \ref{fig:comparison},
\begin{table}
\begin{tabular}{|r|r|r|r|r|r|r|}
\hline
& \multicolumn{3}{c|}{MPI-based} & \multicolumn{3}{c|}{semi-parallel} \\\hline
& flat & rad. & matter & flat & rad. & matter \\\hline
Size & 250 & 180 & 120 & 250 & 180 & 120 \\\hline
Split Factor N &&&& 10 & 10 & 7 \\\hline
Total CPU time (hours) & 28.96 & 47.78 & 73.33 & 8.17 & 32.78 & 14.23 \\\hline
Elapsed time (hours) & 0.887 & 1.60 & 3.75 & 0.155 & 0.669 & 0.595 \\\hline
Total real time (hours) & 56.74 & 102.24 & 240.01 & 9.75 & 37.26 & 17.75 \\\hline
Avg. number of processors & 64.00 & 64.00 & 64.00 & 62.83 & 55.66 & 29.83 \\\hline
CPU utilization percentage & 51.03 & 46.73 & 30.55 & 83.84 & 87.98 & 80.18 \\\hline
Degree of parallelism & 32.66 & 29.91 & 19.56 & 52.67 & 48.97 & 23.92 \\\hline
\end{tabular}
\caption{Comparison of new and old parallelization techniques for
  cosmic string simulation.}
\label{fig:comparison}
\end{table}
we compare the two techniques for several small simulations done in
flat space, in a radiation-dominated universe, and in a
matter-dominated universe.  The flat space case has the most balanced
load, and the matter-dominated case the most unbalanced.  All
simulations were started on 64 processors and the numbers in the table
are the average of 5 runs.  We show the total amount of CPU time used
by all the processors, the elapsed time from start to finish of the
run, and the total amount of processor-time devoted to this task,
whether the processor is running or idle.  In the case of the
MPI-based simulation, all 64 processors are in use simultaneously, so
the total real time is just 64 times the elapsed time.  But in the
semi-parallel simulation, processors with nothing to do are freed, so
that the total real time is a smaller multiple of the elapsed time.
This multiple is the average number of processors, shown in the table.
The CPU utilization percentage is (100 times) the total CPU time
divided by the total real time, and the degree of parallelism is the
total CPU time divided by the elapsed time, i.e., how many times
faster the code ran in the multiprocessing system as compared to a
single processor with 100\% utilization.

In Table \ref{fig:advantages},
\begin{table}
\begin{tabular}{|r|r|r|r|}
\hline
& flat & rad. & matter \\\hline
Size & 250 & 180 & 120 \\\hline
CPU time & 3.54 & 1.46 & 5.15\\\hline
Elapsed time & 5.72 & 2.39 & 6.30 \\\hline
Total real time & 5.82 & 2.74 & 13.52\\\hline
CPU utilization & 1.64 & 1.88 & 2.62\\\hline
Degree of parallelism & 1.61 & 1.64 & 1.22 \\\hline
\end{tabular}
\caption{Advantages of the new parallelization technique over the old.}
\label{fig:advantages}
\end{table}
we show the advantage of the new code over the old.  Each row shows
the ratios of the corresponding quantities in Table
\ref{fig:comparison}.  One can consider the fundamental quantities in
this table to be the CPU time advantage (how much faster the new code
is than the old) and the CPU utilization advantage (how much better
the new parallelization procedure is at keeping the processors doing
useful work).  The total real time advantage is the product of the
advantages in CPU time and CPU utilization; the elapsed time advantage
is the advantage in CPU time multiplied by the improvement in the
degree of parallelism.  Except for the issue of space-time trade-off
described above, the figure of merit of the semi-parallel technique is
the improvement in CPU utilization.  Dividing spacetime volume into
regions with only initial and final surfaces has decreased the amount
of processing resources needed to accomplish the most difficult task
studied here by a factor of 2.6.

In Table \ref{fig:sf},
\begin{table}
\begin{tabular}{|r|r|r|r|} \hline
Split factor & 5 & 6 &  7 \\\hline
CPU time & 15.40 & 14.63 & 14.23 \\\hline
Elapsed time & 0.852 & 0.655 & 0.595 \\\hline
Total real time & 19.28 & 18.12 & 17.75 \\\hline
Avg. number of processors & 22.62 & 27.65 & 29.83 \\\hline
CPU utilization percentage & 79.91 & 80.77 & 80.18 \\\hline
Degree of parallelism & 18.08 & 22.33 & 23.92 \\\hline
\end{tabular}
\caption{The effects of the choice of split factor $N$ for a run of
  size 120 in the matter era.}
\label{fig:sf}
\end{table}
we compare the performance of a small simulation for split
factors $N = 5$, 6 and  7.  The CPU time is somewhat larger for
coarser division of the volume, because we simulate completely each
4-cube that has any point below the final simulation time.  Thus
larger 4-cubes lead to a somewhat larger total volume being
simulated.  The more important effect, however, is that finer division
leads to greater opportunities for parallelism, so that the average
number of processors simultaneously in use and the degree of
parallelism are larger, and consequently the elapsed time is less.

\section{Conclusion}\label{sec:conclusion}

We have implemented a new ``semi-parallel'' technique for
parallelizing a large simulation, as suggested by Pretorius and Lehner
\cite{Pretorius:2003wc}, and used it to do large simulations of cosmic
strings.  We have found this technique to have many advantages, which
we summarize briefly here.

First, because the data does not all need to be processed
simultaneously, we are able to perform larger simulations than would
otherwise be possible.  For example, the simulation described in
Sec.~\ref{sec:strings} used 125,000 regions per layer, and each region
used at maximum about 400 MB of memory.  But a given layer only
occupies 2/3 of the entire volume at most, so the total amount of
memory used for a single time slice through the simulation was 75
TB.\footnote{Of course we have made no attempt to optimize memory
  usage.  On the contrary, we have preferred larger memory usage in
  exchange for a decrease in runtime.}  To our knowledge, only the
largest cluster currently available, TACC Ranger, has enough total
memory to perform such a simulation using conventional techniques.
With our techniques we were able to do this simulation on the Tufts
Research Cluster, which has about 1\% the memory of Ranger.

Furthermore, because the regions can be simulated independently, there
is never any need for processors to wait for each other.  Processor
utilization can be much higher than in conventional parallelism.
For the same reason, there is no need for a particular number of
processors to be available simultaneously.  Since the time to simulate
a single region can be made quite small by choosing small regions,
there is no need to have guaranteed access to processing resources,
and the simulation functions well in environments where jobs can be
preempted.

In the case where one region requires much more effort than its
contemporaries, we may find that all regions not dependent on the
difficult region complete, leaving this region and its future yet to
be done.  In this case, the simulation is, for the moment, no longer
parallel, and the elapsed time until completion may be dominated by
the time to do the job on one processor at a time.  However, it is not
tying up any extra processors.  This is of particular importance when
there are other users of the processor cluster.  The remaining
processors can also be used for additional simulations, so one can
complete many simulations in the same amount of elapsed time that
would otherwise be needed for just one.

Since issues of memory size can be solved by finer division of the
simulation volume, we are free to speed up our code by making
space-time trade-offs in the direction of more space consumption and
lower runtime.

The division into small, independent regions also makes debugging and
dealing with failures much easier.  If a processor crashes, only the
work on its current region is lost.  That region can be started again
from its initial surfaces with little waste of effort.  In contrast, a
conventional parallel simulation would lose all the work done by all
processors since the beginning or since a checkpoint.

Similarly, if the simulation of a region fails due to a bug, that
single region can be run to investigate the problem, without
the need of any other processors.  When the bug is found and
corrected, that region can be restarted and the simulation completed.

This simulation technique could in principle lend itself to ``@Home''
style simulations.  Such a plan is somewhat constrained by the fact
that actual home computers, while often quite fast, are usually
connected to the Internet by fairly slow links, especially for upload
(data transfer from home to the Internet).  So making use of such
computers imposes additional constraints on the ratio of dataset size
to runtime.  It is possible that idle computers at workplaces may be a
better target.  We have not attempted such a project.

\section*{Acknowledgments}

We would like to thank Frans Pretorius, Vitaly Vanchurin and Alex
Vilenkin for helpful conversations.  This work was supported in part
by the National Science Foundation under grant numbers 0855447 and
0903889.  We are grateful for access to the Nemo computing facility at
the University of Wisconsin--Milwaukee where some of these simulations
were performed.  Nemo is funded by part by the National Science
Foundation under grant 0923409.

\bibliographystyle{elsarticle-num}
\bibliography{paper,no-slac}

\end{document}